\documentclass[a4paper,11pt]{article}
\usepackage{amsmath}
\usepackage{epsfig}
\usepackage{graphicx}
\usepackage[round]{natbib}

\title{The Implications of the Early Formation of Life on Earth}
\author{Brendon James Brewer\\
        School of Mathematics and Statistics\\
        The University of New South Wales\\
        \\
        \texttt{brendon.brewer@unsw.edu.au}}
\date{\today}

\begin{document}

\maketitle
\abstract{One of the most interesting unsolved questions in science today is the question of life on other planets. At the present time it is safe to say that we do not have much of an idea as to whether life is common or exceedingly rare in the universe, and this will probably not be solved for certain unless definitive evidence of extraterrestrial life is found in the future. Our presence on Earth is just as consistent with the hypothesis that life is extremely rare as it is with the hypothesis that it is common, since if there was only one planet with intelligent life, we would find ourselves on it. However, we have more information than this, such as the the surprisingly short length of time it took for life to arise on Earth. Previous authors have analysed this information, concluding that it is evidence that the probability of abiogenesis is moderate ($>$ 13\% with 95\% probability) and cannot be extremely small. In this paper I use simple probabilistic model to show that this conclusion was based more on an unintentional assumption than on the data. While the early formation of life on Earth provides some evidence in the direction of life being common, it is far from conclusive, and in particular does not rule out the possibility that abiogenesis has only occurred once in the history of the universe.}

\section{Introduction}
Attempting to make predictions about life elsewhere based on observations about Earth is inherently difficult due to the sample size of 1. It is also fraught with controversial ``anthropic'' considerations \citep{smolin}. However, there is no reason in principle why it cannot be done. If we use probability theory to model uncertainty \citep{jaynes}, and data about life on Earth really is uninformative about extraterrestrial life, then probability theory will return wide probability distributions, indicating the large uncertainty.

The surprising fact that life arose on Earth very quickly after its formation \citep[e.g.][]{early} and at the end of a likely phase of sterilisation due to frequent impacts, has been used to argue that abiogenesis must therefore be easy. \citet[][hereafter L\&D]{2002AsBio...2..293L, 2004IAUS..213..259L} have modelled this reasoning with probability theory and concluded with 95\% confidence (Bayesian posterior probability) that the probability of abiogenesis on an Earth-like planet is greater than 13\%. This was done by using a model where there was constant hazard (chance of life arising per discrete time interval) $q$. The probability distribution for the time $t_L$ (our $t_L$ corresponds to L\&D's $\Delta t_{\textnormal{biogenesis}})$ at which life arises depends on $q$, and this is also calculated conditional on the fact that $t_L$ must be less than the age of the Earth, to correct for the fact that we couldn't have observed the Earth unless life began. This probability distribution is a likelihood function for $q$ when the actual observed $t_L$ is substituted into it. Combined with a prior distribution for $q$, we can then make inferences about its value.

Whilst it is possible and interesting to calculate such things, the model used by L\&D contains a flaw that renders the conclusion invalid. Unfortunately, the conclusion quoted above depends on a choice of prior distribution over $q$ that is overconfident and unrealistic as a description of our state of knowledge about abiogenesis. While uniform priors representing ``initial ignorance'' are common in Bayesian Analysis, a uniform prior for an unknown probability such as $q$ is actually quite informative \citep[][chapter 18]{jaynes}, assigning most of its probability to moderate values of $q$, and ignoring the possibility of extreme values. That the uniform prior is inappropriate can be illustrated using the technique of {\it elaboration}: are we happy with all of the implied consequences of assuming this prior distribution? For instance, one implication is that $q \in [0.49,0.51]$ is just as plausible as $q \in [0, 0.02]$, whereas if we are ignorant, possibilities such as $q \sim 10^{-6}$ and so on should not be ignored as they almost are by the uniform prior. A more realistic representation of complete prior ignorance would be the improper Haldane prior $\propto [q(1-q)]^{-1}$ \citep{jaynes2}, or a modification that removes the singularities at $q=0$ and $q=1$ and makes the prior proper. The Haldane prior corresponds to an improper uniform prior for the ``logit'' $\log[q/(1-q)]$, representing uncertainty not just about the exact value of $q$ but also about its {\it order of magnitude}. This paper uses a model that bypasses direct use of $q$ and deals with expected waiting times instead, although its conclusions can be interpreted in the L\&D framework as well.

The conclusions of \citet{2002AsBio...2..293L} have been criticised previously on the basis that no observer would ever see a recent abiogenesis due to the large number of intermediate steps required between abiogenesis and the development of intelligent life \citep{reply}. However, it is still possible to imagine life arising after 5 Gyr on a planet and intelligent observers discovering this at, say, $t=8$ Gyrs. The fact that we are not in this situation could still be considered surprising, and therefore informative \citep{reply2}.

\section{The Model}
Suppose that there existed a planet that is identical with the early Earth (when conditions have settled down to be suited for life; call this time $t=0$) in terms of all of its macroscopic parameters: mass, temperature, chemical composition, distance from its Sun (which is identical with our Sun), etc. Of course, this model only applies to planets that are Earthlike in terms of their biological characteristics. While this may seem restrictive, observations about what actually occurred on Earth cannot be relevant to planets that do not have this property. Imagine we are given the value of a constant, $\mu$, which is the expected waiting time for the first abiogenesis on a planet with the above conditions. From standard survival analysis, $1/\mu$ is proportional to the probability per unit time of the event happening, and plays the same role as $q$ in L\&D's work. We are then informed, to our great surprise, that the following events occurred on the planet:

- Proposition $S$: At time $t=t_0$ (the present time. Henceforth, a value of 4.3 Gyr is adopted whenever a specific value is required), there exists a person called Brendon James Brewer, and the Prime Minister of Australia on the planet is Kevin Rudd.

- Life first arose on the planet at a time $t_L$. Obviously, $t_L < t_0$.

While proposition $S$ may seem overly specific, one is more likely to make correct inferences by conditioning on a statement that is more specific than, say, ``intelligent life arises''. See \citet{neal} for a detailed discussion of this point and a principled framework for the treatment anthropic selection effects in general.

Our predictions will be given in the form of probability distributions for all of these parameters. The probability distributions are chosen to represent our uncertain state of knowledge - the Bayesian framework \citep{jaynes}. Throughout this paper, probabilities of propositions are denoted by an upper case $P()$ and probability density functions (PDFs) for variables by a lower case $p()$; this notation allows probability expressions to become very succinct as the rules followed by probabilities and PDFs are written in the same way.

\section{Sampling Distribution}
If we only knew the abiogenesis timescale $\mu$, our prediction for $t_L$ would be described by an exponential distribution:
\begin{equation}\label{exponential}
p(t_L|\mu) = \frac{1}{\mu} \exp\left(-t_L/\mu\right) \hspace{2cm} t_L > 0
\end{equation}
Note that this is not an assumption about any frequency distribution that would occur in a population of Earths, it is only the most conservative probability distribution that has the expectation value $\mu$ \citep{jaynes3}. When we find out that $S$ is true for the planet we are watching, the distribution is revised to be truncated to between $t=0$ and $t=t_0$:
\begin{align}\label{truncated}
p(t_L|\mu,S) &= \frac{\frac{1}{\mu} \exp\left(-t_L/\mu\right)}{\int_0^{t_0} \frac{1}{\mu} \exp\left(-t_L/\mu\right) \, dt_L}, \hspace{0.8cm} t_L \in [0,t_0] \\&= \frac{\frac{1}{\mu} \exp\left(-t_L/\mu\right)}{1 - \exp\left(-t_0/\mu\right)}
\end{align}
Technically, this should have been calculated from Bayes' theorem:
\begin{align}
p(t_L|\mu,S) &= \frac{p(t_L|\mu)P(S|t_L,\mu)}{P(S|\mu)} \\
&= \frac{p(t_L|\mu)P(S|t_L,\mu)}{\int_0^\infty p(t_L|\mu)P(S|t_L,\mu) \, d\mu}
\end{align}
where the first term in the numerator would come from Equation~\ref{exponential}. The other term would be very difficult to quantify, however, any effects that they would include apart from the obvious truncation effect ($S$ cannot be true unless $t_L < t_0$) would likely be simply quantitative versions of the evidence and arguments discussed by \citet{reply2}. For example, the fact that it is very unlikely for $S$ to be true if $t_L$ is close to $t_0$ corresponds to the ``non-observability of recent abiogenesis'' and would be modelled in the factor $p(S|t_L,\mu)$. Another possible effect is that there are various epochs in any Earth-like planet's history, and conditions are suitable for life to arise in only one of those epochs. However, for the purposes of this paper, the simple truncation of Equation~\ref{truncated} is sufficient to repeat most of L\&D's argument, while highlighting our point of disagreement with it. Any attempt to increase the sophistication of the model will be deferred to future work.

\begin{figure}
\begin{center}
\includegraphics[scale = 0.5]{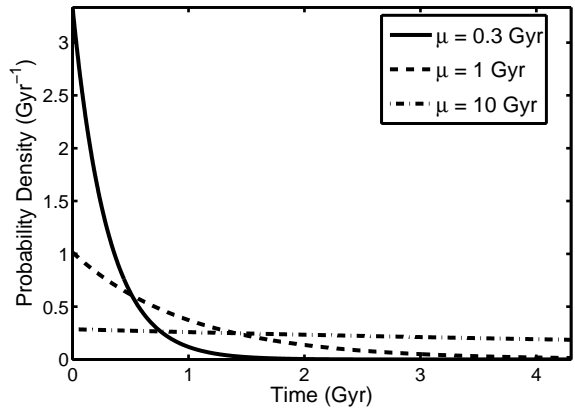}
\caption{The probability density for the time at which abiogenesis occured, given that we exist at t=4.3 Gyrs after the Earth was first suitable for life (defined as t=0). Note that as the abiogenesis timescale becomes larger, this distribution becomes uniform.\label{sampling}}
\end{center}
\end{figure}

The sampling distribution (Equation~\ref{truncated}) for data given parameters is plotted in Figure~\ref{sampling} for three different values of the abiogenesis waiting timescale $\mu$: 0.3, 1 and 10 Gyr. Any proposed value of $\mu$ is a distinct hypothesis that we wish to test, and this sampling distribution defines the predictions that each hypothesis makes about the observational data $t_L$, the actual time that abiogenesis occurred. Note that as $\mu$ increases, this tends to a uniform distribution, and hence {\it moderately large values of $\mu$ ($\sim$ 10 Gyr) and extremely large values of $\mu$ make exactly the same predictions about $t_L$}.

Thus far, this model is virtually identical to that of L\&D - the only difference is that L\&D used a discretised time axis with $\Delta t$=200 Myr and parameterised $\mu$ by $q \approx \mu^{-1} \Delta t$, the probability of life arising in a time $\Delta t$. This makes it difficult to see how they could have extracted such confident conclusions about $q$ based on $t_L$, in light of the above paragraph. This question will be explored in the next section.

\section{Inference About $\mu$}
\begin{figure*}
\begin{center}
\includegraphics[scale = 0.6]{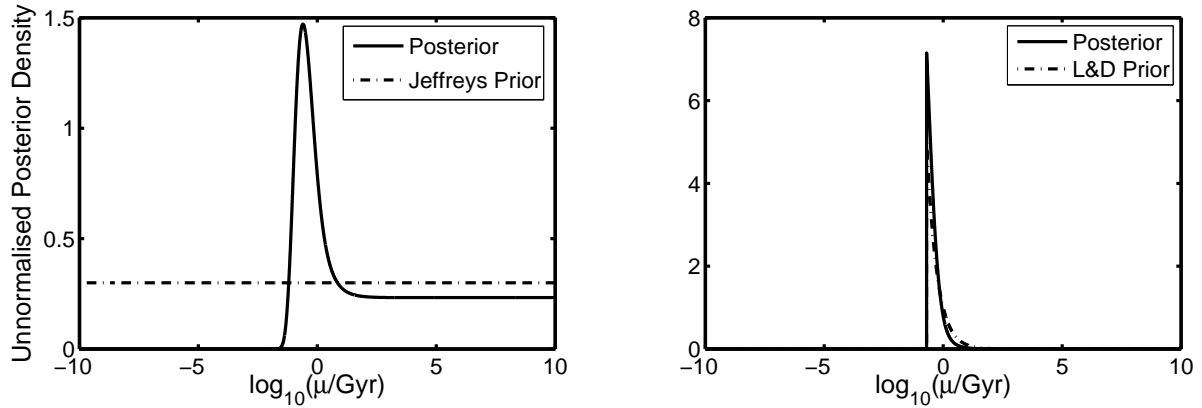}
\caption{The posterior probability density for the logarithm of the abiogenesis timescale, assuming a Jeffreys prior for the timescale (uniform prior for its logarithm), is plotted here as the solid curve in the left panel. The prior that is implied by a uniform prior for the quantity q (the chance of life arising in a finite time interval) is shown as a dotted curve in the right hand panel, along with the resultant posterior. Note that the likelihood function (proportional to our posterior) becomes flat at a nonzero value towards the right of the curve. Hence, whilst the data do support the hypothesis that abiogenesis is likely on Earth-like planets (due to the likelihood peak), it is not a strong enough constraint to rule out more `pessimistic' possibilities.\label{post}}
\end{center}
\end{figure*}

We have a sampling distribution for some data given a parameter of interest, in Equation~\ref{truncated}. To infer the parameter $\mu$ from the known\footnote{$t_L$ is not known exactly, of course. A value of 250 Myr will be adopted whenever a definite value is required.} value of $t_L$, we use Bayes' Theorem to get the posterior distribution for $\mu$, which is proportional to a prior distribution times the likelihood function from Equation~\ref{truncated}:
\begin{equation}
p(\mu|t_L,S) \propto p(\mu|S)p(t_L|\mu,S) = p(\mu)p(t_L|\mu,S)
\end{equation}
Since $S$ by itself hardly tells us anything about any abiogenesis except that it is possible, the dependence on $S$ was dropped from the prior. Now, before we can get probabilistic conclusions about $\mu$ or a related quantity such as $q$, a prior must be assigned. If we are initially ignorant of $\mu$, a suitable prior is the Jeffreys prior $\propto 1/\mu$. The reason for this is that it is equivalent to a uniform improper prior for log($\mu$), and hence describes uncertainty about the {\it order of magnitude} of the parameter. Alternatively, it is the only prior that is invariant under a change of timescale: if we were to find that we are measuring $\mu$ in Terayears rather than Gigayears, the Jeffreys prior is the only choice that would not change in the newly rescaled problem. With this choice, the posterior distribution for $\mu$ cannot be normalised unless we obtain additional information about an upper limit to $\mu$. Hence, it is impossible to construct credible intervals from this data. All we can do is plot the improper posterior for log$_{10}(\mu)$ (which is basically the likelihood, since a Jeffreys prior is uniform for log$_{10}(\mu)$), and this is displayed as the solid curve in Figure~\ref{post}. There is a peak in the posterior, indicating that there is indeed evidence favouring a particular value for $\mu$ of about $t_L$. However, the likelihood flattens out at a non-negative (and non-negligible) value after about $t=$2 Gyr. Thus, this data cannot rule out the hypothesis that $\mu$ is enormous and that Earth hosted the only abiogenesis event(s) in the universe.

This is essentially a quantitative version of an argument that has been put forward previously, \citep[e.g. by][]{hanson}: ``Since no one on Earth would be wondering about the origin of life if Earth did not contain creatures nearly as intelligent as ourselves, the fact that four billion years elapsed before high intelligence appeared on Earth seems compatible with any expected time longer than a few billion years''.

Now, what prior did L\&D implicitly assume? To answer this, a uniform prior for $q$ must be translated to a prior for $\mu$ via the approximate relationship $q \approx \mu^{-1} \Delta t$. Since $q \sim \textnormal{Uniform}(0,1)$, $q/\Delta t = \mu^{-1} \sim \textnormal{Uniform}(0,1/\Delta t)$. By the usual rule for transforming probability distributions:
\begin{align}
p(\mu)\, d\mu &= p(\mu^{-1})\,d(\mu^{-1}) \\
&= p(\mu^{-1}) \frac{d(\mu^{-1})}{d\mu} \, d\mu \\
&= \left(\Delta t\right) \times \left(-\mu^{-2}\right) \, d\mu, \hspace{1cm}\mu \in [\Delta t, \infty]
\end{align}
The negative sign is irrelevant because only the absolute value of the Jacobian matters, so this negative result simply measures the decrease in accumulated probability as one moves leftwards along the $\mu$ axis - so nothing is amiss. It is apparent that the choice of a uniform prior for $q$ is equivalent to a prior for $\mu$ that is proportional to $\mu^{-2}$, and truncated to values of $\mu$ greater than $\Delta t$. This may seem innocuous, but it is significant enough to make the posterior normalisable - in fact, whereas the likelihood function (and posterior wrt a Jeffreys prior) flattens out completely for high $\mu$, the posterior wrt the L\&D prior decays exponentially in that region. The major effect of this choice of prior on the posterior can be seen easily in Figure~\ref{post}.

Unfortunately, unless definitive independent evidence can be found that puts an upper limit on 
possible values of $\mu$, meaningful credible intervals cannot be constructed. L\&D appear to have unwittingly assumed that they did have that extra required information, or that the early formation of life on Earth could provide it, but unfortunately this is not the case. Some information that could provide a likelihood function that allowed the posterior to be normalised would be the following:

- Detection of life elsewhere. Since it is possible to observe a lack of life elsewhere, the sampling distribution of Equation~\ref{truncated} would no longer integrate to 1, and would be truncated at the star's lifetime rather than having anything to do with the age of the Earth. This models some (quite high, in the case of large $\mu$) probability that life will not arise at all on a given planet. It was anthropic considerations that led to the truncation and renormalisation, and these do not apply to the case of life on other planets.

- A very compelling and well understood theory of abiogenesis would enable the direct calculation of $\mu$ from first principles; in theory at least.

These two possibilities, while not exhaustive, would allow definite inferences about $\mu$ that the current data do not. This conclusion accords with the common sense attitute that prevails in the scientific community about what we know and don't know about the probability of abiogenesis.

\section{Conclusion}
The fact that life arose surprisingly early after the formation of the Earth can be used as evidence for the hypothesis that abiogenesis is easy, and hence supports the conclusion that life is common in the universe. However, the evidence is not as conclusive as has been claimed. Specifically, this study has highlighted the fact that knowledge of the early abiogenesis time on Earth is still compatible with the following hypothesis: that life is extraordinarily rare in the universe, perhaps even only on Earth, and we observe early abiogenesis due to chance (we'd have to be moderately lucky, but not obscenely so). This conclusion differs from \citet{2002AsBio...2..293L} because they unwittingly made overconfident prior assumptions. Hence, unless there is a direct detection, the answer to the perennial question ``are we alone'' remains ``nobody knows''.

\section*{Acknowledgments} 
I am supported by an Australian Postgraduate Award and a Denison Merit Award from The School of Physics at The University of Sydney. I would like to thank the anonymous referees of an earlier version of this paper for identifying flaws in the previous version, which allowed the paper to be improved significantly.

\end{document}